\documentclass[doublecol,linenumbers]{epl2}
\usepackage{graphicx}
\usepackage[english]{babel}
\usepackage{latexsym,amsmath,amscd,amssymb,graphicx,amsbsy,color,xcolor,amsfonts,amsthm}

\title{Predicting Mercury's Precession using Simple Relativistic Newtonian Dynamics}
\shorttitle{Predicting Mercury's Precession}
\author{ Y. Friedman \and J. M. Steiner}
\institute{
   Jerusalem College of Technology\\Jerusalem, Israel}

\pacs{95.30.Sf}{Relativity and gravitation}
\pacs{95.10.Eg}{Orbit determination and improvement}

\abstract{
We present a new simple relativistic model for planetary motion describing accurately the anomalous precession of the perihelion of Mercury and its origin. The model is based on transforming Newton's classical equation for planetary motion from absolute to real spacetime influenced by the gravitational potential and introducing the concept of influenced direction.}

\begin{document}
\maketitle

\section{Introduction}

The centennial of the discovery of General Relativity (GR) by Albert Einstein is an opportune moment to strive to fully understand his ideas on gravitation. Although Einstein's GR originates in the logical incompleteness of Newton's Gravitation (NG), that incompleteness itself has not been understood completely, as yet. In our quest to unravel this problem, we incorporated the basic idea of transition from absolute space and time to influenceable spacetime for the problem of planetary motion in NG.

In 1687, Sir Isaac Newton described gravity as an instantaneous and invisible force between two objects. At the end of the 19th century, there seemed to be no dynamic problem which could not be addressed with classical Newtonian mechanics except the anomalous precession of Mercury's orbit.
In 1915, exactly 100 years ago, Albert Einstein devised a completely new description of gravity. In GR, the gravitational force is a fictitious force due to the curvature of spacetime.
This new theory enabled Einstein to calculate the observed anomalous precession of Mercury first recognized in 1859 by Urbain Le Verrier.

In this letter we present a new \textit{simple relativistic} model for planetary motion predicting Mercury's precession without GR. The energy conservation equation for planetary motion in NG is rewritten in terms of dimensionless energy and, then, into a norm equation for the 4-velocity in absolute spacetime. This norm equation is then \textit{transformed} into the corresponding equation in {a spacetime influenced by the gravitational potential - the \textit{real spacetime}}.  Introducing the concept of influenced direction, the resulting equation yields immediately the known equation for planetary motion  predicted by GR. This model predicts the observed value {and also provides an interpretation of Mercury's anomalous precession from the point of view of relativistic Newtonian dynamics.} Finally, we show how to recover the Schwarzschild metric from this equation.

\section{Planetary motion in NG using the concept of dimensionless energy}

We begin with a brief review of planetary motion in NG presented in the literature (see for example \cite{Rindler} - \cite{Brown}) with the introduction of the concept of dimensionless energy.
To describe the motion of a planet of mass $m$ under the gravity of the Sun of mass $M$, we assume that $m\ll M$ and choose a coordinate system with the origin at the center of the Sun. Denote by $\mathbf{r}$ the position vector of the center of the planet relative to the Sun and by $G$ the Newtonian gravity constant.
  The motion of the  planet is governed by Newton's second law
\begin{equation}\label{2Newton}
 m\frac{d^2\mathbf{r}(t)}{dt^2}=-\nabla U(\mathbf{r}(t)),
\end{equation}
where the potential energy of the gravitational field $U=-\frac{GMm}{r}$ and $\nabla U$ denotes the gradient of $U$.
This yields the energy conservation equation in NG
\begin{equation}\label{Energy}
\frac{m}{2}\left(\frac{d\mathbf{r}}{dt}\right)^2-\frac{GMm}{r}=E
\end{equation}
expressing that the total energy $E$ (the sum of the kinetic and potential energies) of the orbit is conserved.

For any energy we define a  \textit{dimensionless energy}  as its ratio  to  the \textit{maximum kinetic energy} (MKE) of the planet $\frac{1}{2}mc^2$, where $c$ is the speed of light. The  \textit{dimensionless kinetic energy} (DKE) is $\frac{1}{c^2}\left(\frac{d\mathbf{r}}{dt}\right)^2=\beta^2$, where $\beta$ is the known beta-factor.
For any point on the orbit, the potential energy is negative and we denote by $u$ the absolute value of the \textit{dimensionless potential energy} (DPE)
\begin{equation}\label{SchwarzSun}
  u=\frac{2GM}{rc^2}=\frac{r_s}{r},\;\;\; r_s=\frac{2GM}{c^2},
\end{equation}
where $r_s$ is the Schwarzschild radius of the Sun  (the radius of a sphere such that, if all the mass  were to be compressed within  a sphere with this radius, the escape velocity  from the surface of the sphere would equal the speed of light).
Finally, we denote by $\mathcal{E}=\frac{2E}{mc^2}$ the \textit{dimensionless total energy} (DTE) of the orbit.

Dividing  equation (\ref{Energy})  by  the MKE of the planet we obtain the conservation equation of the DTE on the orbit
\begin{equation}\label{RE decompNG0}
  \frac{1}{c^2}\left(\frac{d\mathbf{r}}{dt}\right)^2-\frac{r_s}{r}=\mathcal{E}.
\end{equation}

We introduce polar coordinates $r,\varphi$ in the plane of the orbit, where $r$ is the distance of the planet from the Sun and $\varphi$ is the dimensionless polar angle, measured in radians. Conservation of angular momentum per unit mass $J$, allows us to express the angular velocity as $\frac{d\varphi}{dt}=\frac{J}{r^2}$ and to decompose the square of the  velocity of the planet as the sum of the squares of its  orthogonal \textit{radial} and  \textit{transverse} components
 \begin{equation}\label{Theta1_dot}
 \left(\frac{d\mathbf{r}}{dt}\right)^2=\left(\frac{dr}{dt}\right)^2+\frac{J^2}{r^2}.
\end{equation}
 Substituting this into (\ref{RE decompNG0}) we obtain the  classical \textit{dimensionless energy conservation equation}
 \begin{equation}\label{RE decompNG}
 \frac{1}{c^2}\left(\frac{dr}{dt}\right)^2 +\frac{J^2}{c^2r^2}-\frac{r_s}{r}=\mathcal{E}.
\end{equation}

Using the definition (\ref{SchwarzSun}) of $u$, and denoting its derivative with respect to $\varphi$ by $u'$, it can be shown that
\begin{equation}\label{uprime_vel}
\frac{dr}{dt}=-\frac{J}{r_s} u'. \end{equation}
Hence, equation (\ref{RE decompNG}) becomes
\begin{equation}\label{ClassEnergy}
\frac{J^2}{r_s^2 c^2}\left((u')^2+ u^2\right)= u +\mathcal{E}.
\end{equation}
 Multiplying this equation by $2\mu$, where $\mu$ is a\textit{ unit-free orbit parameter}
\begin{equation}\label{zeta}
\mu=\frac{r_s^2 c^2}{2J^2}
\end{equation}
 we obtain
\begin{equation}\label{u_prime}
 (u')^2=-u^2+2\mu u +2\mu \mathcal{E}.
\end{equation}

Differentiating this equation with respect to $\varphi$ and dividing by $2u'$ we obtain a linear differential equation with constant coefficients
\begin{equation}\label{ClasFinal}
  u''+u=\mu.
\end{equation}
Its solution is
\begin{equation}\label{SolU}
  u(\varphi)=\mu(1+\varepsilon\cos (\varphi-\varphi_0)),
\end{equation}
where $\varepsilon$ - the eccentricity of the orbit, and $\varphi_0$ - the polar angle of the perihelion. This implies that
\begin{equation}\label{ClasOrbit}
  r(\varphi)=\frac{r_s/\mu}{1+\varepsilon \cos (\varphi-\varphi_0)}
\end{equation}
and the orbit is a \textit{non-precessing} ellipse. Since the minima of $r(\varphi)$, corresponding to the perihelion, occur when $\varphi=\varphi_0+2\pi n$, $n=0,1,2,\cdots$ the position of the perihelion will not change with the revolution of the planet.

 Moreover, from equation (\ref{SolU}) we obtain that
\begin{equation}\label{mu_meaning}
  \mu=\frac{1}{2\pi}\int_0^{2\pi}u(\varphi)d\varphi
\end{equation}
 is the absolute value of the angular average  DPE on the orbit. The perihelion $r_p$ and aphelion $r_a$ of the orbit correspond to $\frac{dr}{dt}=0$ which by (\ref{uprime_vel}) requires $u'=0$ for their corresponding DPE. Hence, the $u$ values of the perihelion $u_p$  and aphelion $u_a$ are the two (positive) roots of the quadratic on the right hand side of (\ref{u_prime}). This requires
$\mathcal{E}$, the DTE, to be negative for a bound orbit, and from (\ref{SolU})
\begin{equation}\label{elipse_par}
\mu=\frac{u_p+u_a}{2}=\frac{r_s}{L},
\end{equation}
where $L$ is the semi-latus rectum of the  orbit.

The Schwarzschild radius of the Sun is $r_s=2953.25 m$, for Mercury $r_a=6.98168\cdot10^{10}m$ and $r_p=4.60012\cdot10^{10}m$ implying that   $\mu= 5.32497\cdot10^{-8}$ and $\varepsilon=0.20563$ (as  observed). Note that  $\mu\ll 1$.

\section{A new simple relativistic model for planetary motion}

 In the previous section we have seen that NG predicts a non-precessing orbit for planetary motion. Special Relativity (SR) predicts a precession of Mercury, but much lower then the observed one {(see, for example \cite{SIG} p.2033)}. The correct prediction of the observed anomalous precession  is provided by Einstein's GR theory of gravitation based on curved spacetime. In this letter we present a much simpler model which also predicts the correct precession.

 The motion of a planet can be decomposed into two periodic motions:  the radial motion and the angular rotation. In NG, the periods of these  motions are equal, resulting in a non-precessing orbit. Since in reality there is a precession, these two periods are not equal. The reason for this  lies in the inaccurate description of the respective velocities (radial and transverse) of these motions by the NG model.

  Einsten's SR and GR assume that in the classical limit when the velocities  ($DKE\ll 1$) or gravitational fields (or acceleration) ($DPE\ll 1$) are small respectively, the laws of dynamics reduce to the classical Newton's laws. Thus, these theories differ only  in {the way how events} are transformed from a local frame (attached to the moving object) to a fixed inertial reference frame. For instance, (see, for example \cite{F04})  the 3D relativistic dynamics equation in SR can be derived  by transforming Newton's dynamic law from an inertial comoving frame to the reference frame by the use of the Lorentz transformations.

 These spacetime transformations are not trivial. From Planck's formula {(see, for example \cite{Rindler} p.120)} it is evident that time is influenced by energy. SR considers the transformation between inertial frames (with no potential energy). Any rest object in the moving frame has a non-zero kinetic energy in the reference frame. The Relativity Principle \cite{F04}  implies that in this transformation space as well as time  are both influenced by this kinetic energy. The influence on space is in some given direction, which we call the \textit{influenced direction}. By the Lorentz transformations the influenced direction in SR is the direction of the velocity, which entails  that time and the component of the (spatial) displacement vector in this direction are altered by the Lorentz time dilation $\gamma$ factor (depending on DKE), whereas the spatial components transverse to this direction, are not. The same is true for the transformations of the components of the 4-velocity. Since in planetary motion \textit{both}  radial and  transverse components of the velocity are altered, the SR model predicts a precession, but with a value significantly lower than the observed one. Another reason for this low precession is that the influence of the potential energy on spacetime is ignored by the SR model.

We present a new simple relativistic model for planetary motion by transforming the NG solution from an \textit{absolute} (flat) spacetime to the \textit{real} spacetime influenced by the \textit{gravitational potential}, hence by the  DPE. The \textit{gravitational time dilation factor} describing the effect of the DPE on time in such spacetime is (see for example \cite{Rindler})
  \begin{equation}\label{gamma_tilde}
    \tilde{\gamma}=\frac{1}{\sqrt{1-r_s/r}},
\end{equation}
(the analog  of the Lorentz $\gamma$ factor of SR). Note that $\tilde{\gamma}$ has the same form as $\gamma$  with $\beta^2$ replaced  by $r_s/r$, which also reflects the fact that spacetime in SR is influenced by the DKE, while in our model it is influenced by the DPE. It is known that such $\tilde{\gamma}$ yields the true time dilation in the classical limit, and its correctness was verified experimentally in gravitational red-shift experiments.

The gravitational time dilation factor $\tilde{\gamma}$ can be understood (communicated by Z. Weinberger) in terms of the escape velocity $v_e$,  defined as the minimum speed needed for an object to ``break free" from the gravitational attraction of a massive body.  More particularly, it is the velocity (speed travelled away from the starting point) at which the sum of the object's kinetic  and its gravitational potential energies is equal to zero. This is also the velocity of the object pulled by the massive body from infinity. The object keeps record of gravity in the form of the escape velocity, which gets stored in it in the form of (negative) potential energy. For a spherically symmetric massive body, the escape velocity at a given distance is given by
\begin{equation}v_e = \sqrt{\frac{2GM}{r}}.\notag\end{equation}
In terms of $v_e$
\begin{equation}\label{tilde gemma ve}
   \tilde{\gamma}=\frac{1}{\sqrt{1-v_e^2/c^2}},
\end{equation}
which is analogous  to the  $\gamma$ in SR.

Alternatively, we can view  $\tilde{\gamma}$ {as representing} the influence of the acceleration on time. From the definition of the Schwarzschild radius it follows that any real trajectory must satisfy $r>r_s$ in order to prevent it to be absorbed into the Sun. The magnitude of the (classical) acceleration of a free moving object in the gravitational field of the Sun is $a=GM/r^2.$  Hence, by use of (\ref{SchwarzSun}),
\begin{equation}\label{max accel}
  \frac{r_s}{r}<1\;\;\Rightarrow\;\; \frac{r_s}{\sqrt{GM/a}}=\sqrt{\frac{a}{GM/r_s^2}}=\sqrt{\frac{a}{c^2/(2r_s)}}<1 ,
\end{equation}
implying the existence of  a maximal acceleration $a_m$ in our solar system
\begin{equation}\label{mac accel def}
 a_m=\frac{c^2}{2r_s}=1.52164\cdot10^{13}m/s^2
\end{equation}
and hence
\begin{equation}\label{gamm tilde accel}
  \tilde{\gamma}=\frac{1}{\sqrt{1-\sqrt{a/a_m}}}.
\end{equation}
The influence of the acceleration on time is discussed in \cite{F13}  and references therein.

 We  \textit{rewrite} the classical NG dimensionless energy  conservation equation (\ref{RE decompNG}) as
 \begin{equation}\label{RE decompNG2}
 1-\frac{r_s}{r}=-\frac{1}{c^2}\left(\frac{dr}{dt}\right)^2 -\frac{J^2}{c^2r^2}+\mathcal{E}+1.
\end{equation}
 Dividing this equation by the left-hand side we obtain
 \begin{equation}\label{RE decompNG3}
   \frac{1+\mathcal{E}}{1-\frac{r_s}{r}}-\frac{\frac{1}{c^2}\left(\frac{dr}{dt}\right)^2}{1-\frac{r_s}{r}}-\frac{\frac{J^2}{c^2r^2}}{1-\frac{r_s}{r}}=1.
 \end{equation}
 This equation is just another description for the 3D Newtonian orbit with the DPE instead appearing as a separate term, it modifies all the other terms. If we lift this orbit in 4D \textit{absolute} spacetime, this equation resembles a norm equation for the time, radial and transverse components of a 4-velocity, \textit{all}  multiplied by $\tilde{\gamma}^2$. In this case there are only 3 components, because the planar orbit has only two non-zero space components. The multiplication by $\tilde{\gamma}^2$ does not effect the orbit, it merely defines an arc-length parametrization of the classical orbit.

Since the dynamic equation (\ref{2Newton}) involves only the first derivative of the potential $U(r)$ at any given point, it is sufficient to consider the transformation from an absolute to a real spacetime influenced by a \textit{linear} gravitational potential,  the linear part in the expansion of  $U$ at this point. By the Equivalence Principle, such a spacetime is equivalent to a uniformly accelerated system with acceleration $\mathbf{a} =-\nabla U/m$. Using  the Generalised Principle of Relativity, it was shown in \cite{FG10} and \cite{F13}  that, under {this} transformation, both time and the component of velocity in the direction of $\mathbf{a}$ are altered by the same $\tilde{\gamma}$.
Thus, in our proposed model the influenced direction is the \textit{radial} direction. Hence, the transformation from the absolute to the real spacetime should alter (by the time dilation factor $\tilde{\gamma}$) the time and only the radial component of the 4-velocity and not  the transverse ones.

With these ideas we transform the 4-velocity norm  equation (\ref{RE decompNG3}) describing the motion in absolute spacetime to the corresponding norm equation in real spacetime.
 The first term in equation is $\tilde{\gamma}^2,$  the square of the time dilation factor scaled by the DTE of the orbit.
 The second term is the square of the 4-velocity component in the influenced  (radial) direction multiplied by $\tilde{\gamma}^2$ with the arc-length parametrization requiring  that the unit-free radial velocity  $\frac{1}{c^2}\left(\frac{dr}{dt}\right)^2$  be replaced with  $\left(\frac{dr}{ds}\right)^2$. The third term is the square of the transverse component of the 4-velocity also multiplied by $\tilde{\gamma}^2$. However, this term represents a component transverse to the influenced direction (not influenced by the DPE), which should not be affected by our transformation. Hence,the coefficient $\tilde{\gamma}^2$ should be omitted from this term.

This yields  our \textit{modified equation for planetary motion}
\begin{equation}\label{GRnorm}
 \frac{1+\mathcal{E}}{1-r_s/r}-\frac{\left(\frac{dr}{ds}\right)^2}{1-r_s/r}-\frac{J^2}{c^2r^2}=1
\end{equation}
implying
\begin{equation}\label{r_dotGR}
  \left(\frac{dr}{ds}\right)^2+\frac{J^2}{c^2r^2}\left(1-\frac{r_s}{r}\right)-\frac{r_s}{r}=\mathcal{E}.
\end{equation}

A close inspection reveals that this equation is simply a minor modification of the NG dimensionless energy conservation equation (\ref{RE decompNG}) in which \textit{only} the transverse component of the DKE is multiplied by $\tilde{\gamma}^{-2}$.
This equation is analogous to the known GR equation for planetary motion, as the geodesic of the Schwarzschild metric.

Our model also reveals the source of the precession of the planetary orbit. As mentioned above, in NG, the radial and the transverse periods are identical, resulting in a non-precessing orbit. In SR, \textit{both} the radial and transverse components of the velocity are altered, resulting in unequal periods with relatively small difference between them and hence a small precession. In our  model, \textit{only} the radial component of the velocity is influenced, while the transverse (angular) component is not. This, in turn,  accentuates the difference between these periods, resulting in the  observed precession, as follows.

To define the precise value of the precession, we rewrite
equation (\ref{r_dotGR}) as
\begin{equation}\frac{J^2}{c^2r_s^2}(u')^2=\frac{J^2}{c^2r_s^2}u^3-\frac{J^2}{c^2r_s^2}u^2+u +\mathcal{E}.\notag\end{equation}
 Multiplying this equation by $2\mu,$  with $\mu$ defined by (\ref{zeta}) we obtain
\begin{equation}\label{u_primeGR}
 ( u')^2=u^3-u^2+2\mu u +2\mu \mathcal{E}.
\end{equation}
This equation is identical to (\ref{u_prime}) in {NG, except that it has a} very small (since $u\ll 1$) additional term $u^3$ on the right-hand side. {This result is the same result as that of GR.}

We seek a solution of this equation in the form generalizing (\ref{SolU}),
\begin{equation}\label{GRu_form}
  u(\varphi)=\mu(1+\varepsilon\cos\alpha(\varphi))
\end{equation}
for some function $\alpha(\varphi)$. As before, two roots of the cubic on the right-hand side of (\ref{u_primeGR}), are the $u$ values  $u_p$ and $u_a$ of the the perihelion and aphelion, respectively.  Moreover, since the coefficients of this cubic are constant for a given orbit, these values will not change from one revolution to the next. We denote the third root of this cubic by $u_e$. Thus, equation (\ref{u_primeGR}) can be factorized as
\begin{equation}\label{u_primeGRprod}
 ( u')^2=-(u-u_p)(u-u_a)(u_e-u).
\end{equation}

 From equation (\ref{GRu_form}), $( u')^2=(\alpha')^2\mu^2\varepsilon^2\sin^2\alpha(\phi)$, $u_p=\mu+\mu\varepsilon$ and $u_a=\mu-\mu\varepsilon.$
 Moreover, since the sum of the roots of this cubic is 1, \begin{equation}u_e=1-(u_p+u_a)=1-2\mu.\end{equation}
Substituting  these into (\ref{u_primeGRprod}), yields after simplification
 \begin{equation}\alpha'=\frac{d\alpha}{d\varphi} =(1-3\mu-\mu\varepsilon\cos\alpha(\varphi))^{1/2}.\notag\end{equation}
 This allows us to obtain the dependence of $\varphi$ on $\alpha$ as
 \begin{equation}\label{phy_alpha}
   \varphi(\alpha) =\varphi_0+\int_0^\alpha (1-3\mu-\mu\varepsilon\cos\tilde{\alpha})^{-1/2}d\tilde{\alpha}.
 \end{equation}

 As mentioned above, for  Mercury, $\mu$ is very small.  Expanding the integrand into a power series in $\mu,$ we obtain
 \begin{equation}\varphi(\alpha) =\varphi_0 +\alpha +\frac{3}{2}\mu \alpha +\frac{\mu}{2}\varepsilon\sin \alpha+\cdot\cdot\cdot\,.\notag\end{equation}
 The polar angles of the  perihelion correspond to $\alpha= 2\pi n, \;n=0,1,2,\cdots $.  Thus, the precession of the perihelion, up to the first order in $\mu$, is  given by
 \begin{equation}\label{precesion}
  \varphi(2\pi)-\varphi(0)-2\pi\approx 3\pi\mu.
 \end{equation}
 From this equation, the precession of the perihelion of Mercury is  $5.01866\cdot10^{-7}$ radians per revolution, which is \textit{exactly} the currently observed one.

Finally, we show how we can recover the Schwarzschild metric from equation (\ref{GRnorm}). Since the influenced direction in our model is the radial direction, our metric will differ from the Minkowski metric only in the time and radial components. Thus, our metric in spherical coordinates is
\begin{equation}ds^2=f(r)dx^0-g(r)(dr)^2-r^2((d\theta)^2+\cos ^2\theta (d\varphi)^2).\end{equation}
The orbit is  the geodesic of this metric optimizing the Lagrangian $L(x^0,r, \varphi, \theta,\dot{x}^0,\dot{r}, \dot{\varphi}, \dot{\theta})$,
\begin{equation}L= f(r)(\dot{x}^0)^2-g(r)(\dot{r})^2
- r^2((\dot{\theta})^2+\cos ^2\theta (\dot{\varphi})^2)\notag\end{equation}
where the \textit{dot} denotes  differentiation with respect to $s$.

Assuming $\theta =\pi/2$, the Euler-Lagrange equations yield  $f(r)\dot{x}^0=a$ for some constant $a$ and $r^2\dot{\varphi}=J/c$. The norm of the 4-velocity on the orbit is
\begin{equation} f(r)(\dot{x}^0)^2-g(r)(\dot{r})^2 - r^2 (\dot{\varphi})^2=\frac{a^2}{f(r)}-g(r)(\dot{r})^2 - \frac{J^2}{c^2r^2}.\notag\end{equation}
 Comparing this with (\ref{GRnorm}) we obtain  $f(r)=1-r_s/r$ and $g(r)=\frac{1}{1-r_s/r},$ implying that the metric is the Schwarzschild metric.

\section{Discussion}

Our model for planetary motion presented here is a special case  of a more general  Newtonian relativistic dynamics, which will be presented in \cite{F16}. In this dynamics, the energy conservation equation for motion under a conservative force  in  a spacetime influenced by the potential energy $U(\mathbf{x})$  (vanishing at infinity) is
\begin{equation}\label{NRD}
  \frac{m}{2}\left( (1-u)\dot{\mathbf{x}}^2+u(\dot{\mathbf{x}}\cdot\mathbf{n})^2\right) + U(\mathbf{x})=E,
\end{equation}
\vspace{12pt}
where $u=-\frac{2U}{mc^2},\;\mathbf{n}=\frac{\nabla U}{|\nabla U|}$ and \textit{dot} denotes differentiation with respect to time.

In the case of a radial potential $U(r)$, $\mathbf{n}$ is in the radial direction and the  conservation of angular momentum defines explicitly the velocity decomposition  into its radial and transverse components. Equation (\ref{NRD}) yields
\begin{equation}\label{RadNRD}
 \left(\frac{dr}{dt}\right)^2=-\frac{J^2}{r^2}(1-u)-\frac{2U}{m}+\frac{2E}{m},
\end{equation}
which reduces to (\ref{r_dotGR}) for planetary motion.

Our model considers only the influence of the gravitational potential and ignores the influence of the kinetic energy. A complete model should consider both these effects. Nevertheless, our model correctly predicts Mercury's anomalous precession  either because the influence of the kinetic energy is below the experimental accuracy or, its effect is cancelled  out by the interaction with the neighbouring planets.

\acknowledgments We wish to acknowledge Prof. Lawrence Paul Horwitz and Mr. Zvi Weinberger for their constructive comments.

\end{document}